\PassOptionsToPackage{x11names}{xcolor}

\documentclass[
    aps,
    prd,
    onecolumn,
    tightenlines,
    notitlepage,
    superscriptaddress,
    nofootinbib,
    preprintnumbers,
    floatfix,
    showkeys,
    11pt,
    altaffilletter
]{revtex4-2}

\usepackage[T1]{fontenc}
\usepackage[utf8]{inputenc} 
\usepackage{lmodern}

\usepackage{amsmath,amssymb,amstext,amsfonts}
\usepackage{mathrsfs}
\usepackage{slashed}
\usepackage{cancel}
\usepackage{upgreek}
\usepackage{nccmath}
\allowdisplaybreaks

\usepackage[version=4]{mhchem}
\usepackage[official]{eurosym}
\usepackage{textgreek}
\usepackage{textcomp}
\usepackage{gensymb}

\usepackage{graphicx}
\graphicspath{{figures/}}

\usepackage{float}
\usepackage{caption}
\usepackage{subcaption}
\usepackage{booktabs}
\usepackage{multirow}
\usepackage{adjustbox}

\usepackage[normalem]{ulem}
\usepackage{enumerate}
\usepackage{comment}

\usepackage{url}
\usepackage{orcidlink}
\usepackage{fontawesome}

\usepackage{xcolor}
\definecolor{cerulean}{rgb}{0.0, 0.48, 0.65}
\usepackage{hyperref}
\usepackage[capitalise]{cleveref}
\hypersetup{colorlinks,citecolor=cerulean,linkcolor= cerulean}


\AtBeginDocument{\hypersetup{citecolor=cerulean,linkcolor=cerulean,urlcolor=cerulean}}

\def\cevns{CE\textnu NS}

\def\d{\mathrm{d}}
\newcommand{\qtransfer}{\left|\mathbf{q}\right|}

\definecolor{amber}{rgb}{1.0, 0.49, 0.0}

\begin{document}
\sloppy

\title{{\Large Refined extraction of electroweak and nuclear parameters\\ from germanium \cevns~data}}

\author{Valentina De Romeri~\orcidlink{0000-0003-3585-7437}}
\email{deromeri@ific.uv.es}
\affiliation{Instituto de F\'{i}sica Corpuscular (IFIC), CSIC‐Universitat de Val\'encia, E-46980 Valencia, Spain}

\author{Laura Duque~\orcidlink{0009-0009-6109-7946}}
\email{laura.duque@cinvestav.mx}
\affiliation{Departamento de F\'{\i}sica, Centro de Investigaci\'on
  y de Estudios Avanzados del IPN,\\ Apartado Postal 14-740 07000 Ciudad de Mexico, Mexico}

\author{Dimitrios K. Papoulias~\orcidlink{0000-0003-0453-8492}}\email{dimitrios.papoulias@uni-hamburg.de}
\affiliation{Institute of Experimental Physics, University of Hamburg, 22761, Hamburg, Germany}

\author{G. Sanchez Garcia~\orcidlink{0000-0003-1830-2325}}%
\email{g.sanchez@ciencias.unam.mx}%
\affiliation{Departamento de F\'{i}sica, Facultad de Ciencias, Universidad Nacional Aut\'onoma de M\'exico,
Apartado Postal 70-542, Ciudad de M\'exico 04510, M\'exico}%

\author{Christoph A. Ternes~\orcidlink{0000-0002-7190-1581}}
\email{christoph.ternes@lngs.infn.it}
\affiliation{Gran Sasso Science Institute, Viale F. Crispi 7, L’Aquila, 67100, Italy}
\affiliation{Istituto Nazionale di Fisica Nucleare (INFN), Laboratori Nazionali del Gran Sasso, 67100 Assergi, L’Aquila (AQ), Italy
}


\begin{abstract}
We present a combined analysis of recent \cevns~data on germanium from two complementary experiments: COHERENT, which uses neutrinos from pion decay  at rest, and CONUS+, which detects reactor antineutrinos. Exploiting the complementarity of these two datasets in a joint statistical analysis, we extract the germanium root-mean-square neutron radius and neutron skin with improved precision, disentangling spectral shape distortions from overall normalizations and reducing systematic uncertainties. 
We also determine the  weak mixing angle at low momentum transfer, providing a test of the Standard Model in a less-explored kinematic regime. A key systematic uncertainty in \cevns~ionization measurements is the nuclear quenching factor; we therefore present our results as a function of variations of the Lindhard model. 
For the nuclear form factor, we adopt the analytical  Klein-Nystrand parametrization and benchmark it against predictions from the large-scale nuclear Shell Model, assessing the impact of nuclear structure uncertainties on our results.
Our analysis demonstrates the power of combining datasets across different neutrino sources to maximize sensitivity to both nuclear and electroweak physics.

\end{abstract}
\maketitle

\section{Introduction}

Low-energy neutrino detection via neutral-current interactions has emerged as a powerful probe, complementary to conventional charged-current channels. In this context, coherent elastic neutrino-nucleus scattering (\cevns) is particularly relevant due to its coherent enhancement, where the neutrino interacts with the nucleus as a whole~\cite{Freedman:1973yd}. Although experimentally challenging for decades because of the tiny nuclear recoil energies, recent advances have enabled a rapidly growing \cevns~program, providing robust tests of the Standard Model (SM) and sensitivity to new physics (see, for instance, Refs.~\cite{Barranco:2005yy,Lindner:2016wff,Billard:2018jnl,Papoulias:2019xaw,Papoulias:2019lfi,AristizabalSierra:2019ufd,Cadeddu:2020lky,Galindo-Uribarri:2020huw,Miranda:2020tif,Coloma:2020nhf,Cadeddu:2021ijh,DeRomeri:2022twg,Coloma:2022avw,Sierra:2023pnf,Rossi:2023brv,Liao:2024qoe,AtzoriCorona:2025xwr}). 

A variety of artificial and natural neutrino sources have been set to measure \cevns~\cite{Abdullah:2022zue}. The first observation was reported by the COHERENT collaboration using pion decay at rest neutrinos at the Spallation Neutron Source (SNS) and a CsI target~\cite{COHERENT:2017ipa}. 
Since then, measurements with CsI~\cite{COHERENT:2021pcd}, Ar~\cite{COHERENT:2020iec}, and Ge~\cite{COHERENT:2024axu,COHERENT:2026yje} detectors at the same facility have probed the dependence of the SM cross section on the nuclear neutron number. 
Reactor experiments such as Dresden-II~\cite{Colaresi:2022obx} and CONUS+~\cite{Ackermann:2025obx} have reported complementary measurements with electron antineutrinos, also on Ge detectors, while evidence for solar $^8$B-induced \cevns~has been observed in Xe-based dark matter detectors~\cite{PandaX:2024muv,XENON:2024ijk,LZ:2025igz,XENON:2026ydt}.

In this work, we focus on the recent high-precision COHERENT measurement on a Ge target~\cite{COHERENT:2026yje}, obtained with approximately three times the exposure of their first Ge result~\cite{COHERENT:2024axu} and improved analysis techniques enabling a lower threshold, translating into six times more signal statistics compared to the previous result. Due to the typical energies of a few tens of MeV at the SNS, this dataset probes neutrinos in a regime where coherence is partially lost, enhancing sensitivity to nuclear form factors and finite-size effects. 
This stands in contrast to reactor neutrino experiments such as CONUS+, where energies are below 10 MeV and the interaction occurs deep in the fully coherent regime. While the sensitivity to nuclear-structure effects is therefore reduced, the much lower detector threshold achieved at CONUS+ provides access to the low-energy recoil spectrum, making it particularly sensitive to detector-response effects such as quenching. We exploit this complementarity between the two experiments~\cite{COHERENT:2026yje,Ackermann:2025obx} to improve constraints on the neutron root-mean-square (rms) radius of germanium, $\langle R_n\rangle$, and to extract a precise determination of the weak mixing angle, $\sin^2\theta_W$, at low momentum transfer.

A non-trivial systematic uncertainty in ionization-based detectors used for \cevns, is the nuclear quenching factor, which relates the true nuclear recoil energy to the measured ionization signal. The low-energy behavior of this quantity remains under debate~\cite{Collar:2021fcl,Bonhomme:2022lcz}, particularly regarding the validity of the Lindhard model~\cite{osti_4536390} and possible deviations from it.
As the phenomenological implications of \cevns~measurements have been shown to depend strongly on the choice of quenching factor~\cite{Li:2025pfw}, we treat the Lindhard parameter, $k$, as well as the parameters $\langle R_n\rangle$ and  $\sin^2\theta_W$ as simultaneous free parameters, performing a joint analysis to map out their correlations and assess the impact of quenching uncertainties on our results. 
We further include momentum-dependent radiative corrections to neutrino couplings following Ref.~\cite{AtzoriCorona:2024rtv}, which are relevant for a precise determination of $\sin^2\theta_W$. Finally, we assess theoretical uncertainties from nuclear structure by comparing two descriptions of the nuclear form factor: the Klein-Nystrand parametrization~\cite{Klein:1999qj} and calculations based on the large-scale nuclear Shell Model~\cite{Hoferichter:2020osn}.

The paper is organized as follows. In Sec.~\ref{sec:theory}, we introduce the theoretical framework underlying the \cevns~cross section and the calculation of the predicted event rates. Section~\ref{sec:analyses} summarizes the experimental inputs and statistical methodology adopted in the analyses of the COHERENT-Ge and CONUS+ datasets. Our results are presented in Sec.~\ref{sec:results}, while Sec.~\ref{sec:concl} contains our conclusions.

\section{CE$\nu$NS~theory and signal prediction}
\label{sec:theory}

In this section, we summarize the theoretical framework adopted for the description of \cevns, including the relevant cross sections and nuclear form factors.

\subsection{\cevns~cross section}

Within the SM, the differential \cevns~cross section with respect to the nuclear recoil energy $T_\mathcal{N}$ is given by~\cite{Freedman:1973yd}
\begin{equation}
\label{eq:xsec_CEvNS}
\frac{\d\sigma_{\nu \mathcal{N}}}{\d T_\mathcal{N}}\Big|_\mathrm{CE\nu NS}=\frac{G_F^2 m_\mathcal{N}}{\pi}\left({Q_V^\mathrm{SM}}\right)^2\left(1-\frac{m_\mathcal{N} T_\mathcal{N}}{2E_\nu^2}\right) \, ,
\end{equation}
where $G_F$ is the Fermi constant, $E_\nu$ represents the incident neutrino energy, and $m_\mathcal{N}$ denotes the nuclear mass. The SM vector weak charge, $Q_V^\text{SM}$, is defined as
\begin{equation}
\label{eq:CEvNS_SM_Qw}
    Q_V^\text{SM} = g_V^p Z  F_p(\qtransfer)+ g_V^n N  F_n(\qtransfer)\, ,
\end{equation}
with $Z$ and $N$ the number of protons and neutrons in the nucleus, respectively. At tree level, the vector couplings in the previous equation are flavor independent and given by $g_V^p = (1- 4 \sin^2 \theta_W)/2$ and $ g_V^n = -1/2$.  In our analysis, we include the momentum-dependent radiative corrections to these couplings following Ref.~\cite{AtzoriCorona:2024rtv,AtzoriCorona:2025xwr}, which induce a mild flavor dependence\footnote{Although the momentum dependence of $g_V^p$ has a negligible numerical impact in the \cevns~energy regime, it is consistently included in our analysis.} (see also Refs.~\cite{Tomalak:2020zfh,Tomalak:2021lif,Tomalak:2025tls} for other works addressing radiative corrections relevant for \cevns). 

Nuclear-physics effects are incorporated through the nuclear proton ($F_p$) and neutron ($F_n$) form factors. These effects are especially relevant for the COHERENT Ge measurement, where the momentum transfer begins to probe the loss of full coherence. As our main approach, we adopt the phenomenological Klein-Nystrand (KN) parametrization~\cite{Klein:1999qj}
\begin{equation}
  \label{eq:KNFF}
   F_{p,n}(\qtransfer^2)=3\frac{j_1(\qtransfer R_A)}{\qtransfer R_A} \left(\frac{1}{1+\qtransfer^2a_k^2} \right)\, ,
\end{equation}
where $\qtransfer = \sqrt{2m_\mathcal{N}T_\mathcal{N}}$ is the three-momentum transfer, $j_1$ denotes the spherical Bessel function of first order, and $a_k=0.7~\mathrm{fm}$ is the Yukawa diffuseness parameter. The diffraction radius, $R_A$, is related to the proton or neutron root-mean-square (rms) radius through
\begin{equation}
\label{eq:rms}
   R_A^2 = \frac{5}{3} \left(\left<R_{p,n}\right>^2 -6a_k^2 \right) \, .
\end{equation}
For the germanium isotopes relevant to this work, we fix the proton rms radius to $\langle R_p\rangle = 4.078~\mathrm{fm}$, corresponding to the average value for all isotopes~\cite{Wang:2024ste,Angeli:2013epw}.  In contrast, the neutron rms radius remains poorly constrained and is commonly expressed in terms of the neutron skin 
\begin{equation}
\Delta_{np} = \langle R_n\rangle - \langle R_p\rangle.
\end{equation}
For completeness, we stress that within the phenomenological treatment adopted for the nuclear form factors, the neutron skin extracted from \cevns~data relies on identifying the proton rms radius entering the weak form factor with the experimentally measured charge radius. For germanium isotopes, this approximation induces a shift of approximately $0.1~\mathrm{fm}$ relative to the corresponding point-density neutron skin definition. Such an effect remains subleading compared to the current experimental uncertainties; see Refs.~\cite{Coloma:2020nhf,Sierra:2023pnf} for further discussion.

We also assess the robustness of the phenomenological KN form factor by comparing its predictions against calculations based on the full nuclear response. For this purpose, we employ the multipole expansion of the hadronic current developed by Donnelly and Walecka~\cite{Donnelly:1976fs}. In this framework, vector nuclear responses are computed within the Shell Model, with  the  relevant contributions  for \cevns~ arising from the Coulomb ($\mathcal{M}$) and $\Phi''$ operators as defined in Appendix B of Ref.~\cite{Hoferichter:2020osn}. We compute the full vector nuclear response through the modified weak nuclear form factor defined in Eq.~(52) of~\cite{Hoferichter:2020osn}, including the radiative corrections to $g_V^p$ and $g_V^n$ discussed above\footnote{Note that our convention for the SM couplings differs by a factor of two from that adopted in Ref.~\cite{Hoferichter:2020osn}.}.
Notice that the $\Phi''$ contribution is suppressed by powers of $\qtransfer$, while the Coulomb response contains subleading nuclear corrections that are likewise $\qtransfer$ suppressed and can be safely neglected in the rms-radius definition of Eq.~(\ref{eq:rms}).

We conclude noticing that since the \cevns~cross section actually probes $(Q_V^\mathrm{SM})^2 \propto F_n^2(\qtransfer^2)$, with the form factor depending on the product $\qtransfer \times \langle R_n\rangle$, higher-energy neutrinos --- corresponding to larger momentum transfer --- provide enhanced sensitivity to the form factor suppression and thus to the neutron radius. In this respect, the combination with reactor data, which mainly constrain the overall normalization in the fully coherent regime, is expected to contribute to  disentangle spectral distortions induced by nuclear-structure effects.

\subsection{Event rates}
\label{subsec:rates}

We now discuss the computation of the predicted event rates to be compared with \cevns~data from the COHERENT and CONUS+ experiments. In both cases, we first evaluate the differential event rate as
\begin{eqnarray}
\label{eq:dRdEr_CEvNS}
    \left. \dfrac{\d R}{\d E_\text{reco}}\right|_{\rm CE\nu NS} &= \mathcal{E} \int_{E_\mathrm{er}^\mathrm{min}}^{E_\mathrm{er}^\mathrm{max}} \d E_\text{er} \, \mathcal{G}(E_{\rm reco},E_\text{er}) \mathcal{A}(E_\text{er})\int_{E_\nu^\mathrm{min}}^{E_\nu^\mathrm{max}}  \, \d E_\nu \, \,  \dfrac{\d \phi}{\d E_\nu} \left. \dfrac{\d \sigma_{\nu \mathcal{N}}}{\d E_\text{er}}\right|_{\rm CE\nu NS} \,,
\end{eqnarray}
where we distinguish between the reconstructed ionization energy $E_{\rm reco}$ and the true ionization energy $E_{\rm er}$, with the subscript indicating units of electron recoil. Hence, we need the differential cross section in terms of $E_{\rm er}$, obtained via
\begin{equation}
 \left.\dfrac{\d \sigma_{\nu \mathcal{N}}}{\d E_\text{er}}\right|_{\rm CE\nu NS} = \frac{\d T_\mathcal{N}}{\d E_\text{er}}
   \left.\dfrac{\d \sigma_{\nu \mathcal{N}}}{\d T_\mathcal{N}}\right|_{\rm CE\nu NS} = \frac{1}{\mathrm{Q_F}} \left( 1- \frac{E_\text{er}}{\mathrm{Q_F}} 
   \frac{\d \mathrm{Q_F}}{\d E_\text{er}} \right) \left.\dfrac{\d \sigma_{\nu \mathcal{N}}}{\d T_\mathcal{N}}\right|_{\rm CE\nu NS}, 
\label{eq:CEvNS_change_variable}
\end{equation} 
where $\textrm{Q}_\textrm{F}$ is the quenching factor,   which accounts for the fraction of energy lost into heat during ionization. As discussed in the Introduction, we adopt the Lindhard Model~\cite{osti_4701226,PhysRev.124.128} 
\begin{equation}
    E_{\rm er} = \mathrm{Q_F}(T_\mathcal{N}) \times T_\mathcal{N} = \frac{k~g(\epsilon)}{1+ k~g(\epsilon)}T_\mathcal{N} \, , 
\label{eq:quenching}    
\end{equation}
with  $\epsilon = 11.5~Z^{-7/3} T_\mathcal{N}$, and $g(\epsilon) = 3 \epsilon^{0.15} + 0.7 \epsilon^{0.6} + \epsilon$.
Coming back to Eq.\eqref{eq:dRdEr_CEvNS}, the integration limits follow from the kinematics of the process, with $E_\mathrm{er}^\mathrm{min}=2.98~\mathrm{eV_{ee}}$ corresponding to the minimum  energy required to produce a hole-pair in germanium~\cite{Coloma:2022avw}. The remaining factors are detector-dependent and are briefly summarized here, with further details provided in the next section. In Eq.~\eqref{eq:dRdEr_CEvNS}, $\mathcal{E}$ parametrizes the exposure, $\mathcal{A}(E_{\rm er})$ the detection efficiency (taken to be close to unity in both experiments), and $\d \phi/\d E_\nu$ the neutrino flux. For COHERENT, we consider neutrinos produced via pion and muon decays at rest at the Spallation Neutron Source~\cite{COHERENT:2021yvp}, while for CONUS+ we adopt the reactor antineutrino flux parametrization provided in the Appendix of Ref.~\cite{CONNIE:2019xid}.
We account for detector energy resolution via a Gaussian smearing function characterized by the energy resolution
\begin{equation}
    \sigma(E_{\textrm{er}}) = \sqrt{\sigma_{\textrm{noise}}^2 + \eta \times \mathcal{F} \times E_{\textrm{er}} }\, ,
\end{equation}
where $\eta = 2.98~\mathrm{eV_{ee}}$, $\sigma_{\textrm{noise}}$ accounts for electronic noise, and $\mathcal{F}$ is the Fano factor which will be specified below.

We further account for the natural isotopic composition of germanium\footnote{We have checked that using an average mass number $A=72.6$ leads to negligible changes in the predicted signal.}, namely $^{70}$Ge (20.57\%), $^{72}$Ge (27.45\%), $^{73}$Ge (7.75\%), $^{74}$Ge (36.50\%), and $^{76}$Ge (7.73\%).
Finally, our reconstructed event spectra,  $R^{\rm th}_i$ for the $i$-th energy bin, are computed through the integral
\begin{equation}
    R_i = \int_i  \dfrac{\d R}{\d E_{\rm reco}} \d E_{\rm reco}\, ,
\label{eq:ev_rate}    
\end{equation}
where we have considered 28 energy bins in $(0.5,20)~\mathrm{keV_{ee}}$ for COHERENT-Ge with bin widths as given in Ref.~\cite{COHERENT:2026yje}, and 19 evenly-spaced energy bins in $(0.16,3.5)~\mathrm{keV_{ee}}$  for CONUS+~\cite{Ackermann:2025obx}.

\section{Data analyses}
\label{sec:analyses}

In this section, we present the experimental inputs and describe the statistical procedure adopted in the analysis of the COHERENT-Ge and CONUS+ datasets.

\subsection{COHERENT Ge-Mini}
\label{subsec:COH-Ge}

The COHERENT experiment has performed two data-taking campaigns with Ge detectors: the first between June and August 2023~\cite{COHERENT:2024axu}, and the second between February and May 2025~\cite{COHERENT:2026yje}. Given the higher exposure and improved statistics, we focus on the latter campaign in this work. 

The Ge-Mini detector of COHERENT consists of an
array of p-type point contact germanium detectors, and is located at a baseline of $L=19.2$~m from the SNS. The measurement considered here was performed with an array of four detectors with a total  fiducial mass of 8.53~kg. 
The SNS neutrino flux is normalized as $\Phi = r N_{\mathrm{POT}}/(4\pi L^2)$, where $N_{\mathrm{POT}} = 4.68\times10^{22}$ is the total number of protons on target during the data-taking period and $r=0.37$ is the neutrino yield per POT~\cite{COHERENT:2026yje}. For the detector response and energy reconstruction, we adopt\footnote{Since these quantities are not explicitly provided in the official release, we use the values reported in Ref.~\cite{Bouabid:2025upo}, obtained as averages over the different sub-detectors.} an electronic noise of $\sigma_{\mathrm{noise}} = 58.17~\mathrm{eV_{ee}}$ and a Fano factor $\mathcal{F}=0.066$.

For the statistical analysis, we construct a Poissonian $\chi^2$ function
\begin{equation}
  \chi^2(\vec{{S}}) = 2\sum_{i}\left [ R^{\textrm{th}}_{i}(\vec{S}) - R^{\textrm{exp}}_{i} + R^{\textrm{exp}}_{i}\ln\left ( \frac{R^{\textrm{exp}}_{i}}{R^{\textrm{th}}_{i}(\vec{S})} \right )\right ] + \left(\frac{\alpha}{\sigma_{\alpha}} \right)^2+ \left( \frac{\beta}{\sigma_{\beta}} \right)^2 \, ,
    \label{eq:chi:future}
\end{equation}
where $\vec{S}=\{\sin^2\theta_W,\langle R_n\rangle,k\}$ denotes the set of fitted SM parameters. The theoretical prediction in each bin is given by
\begin{equation}
    R^{\textrm{th}}_i(\vec{S}) = (1+\alpha) R^{\textrm{CE$\nu$NS}}_i(\vec{S}) + (1+\beta)R^{\textrm{SSB}}_i \, ,
    \label{N:chi:future}
\end{equation}
where $ R^{\textrm{CE$\nu$NS}}_i(\vec{S})$ is estimated from Eq.~\eqref{eq:ev_rate} and  $R^{\textrm{SSB}}_i$ represents the steady state background (SSB) taken from Ref.~\cite{COHERENT:2026yje}.
Systematic uncertainties are incorporated via two nuisance parameters. The parameter $\alpha$ accounts for signal systematics arising from: the neutrino flux (10\%)~\cite{COHERENT:2026yje}, detector baseline (0.5\%), energy calibration (0.8\%), active mass (2\%), nuclear form factor (0.8\%), and quenching factor (0.7\%), leading to an overall normalization uncertainty of $\sigma_\alpha = 10.3\%$. The parameter $\beta$ describes the SSB normalization uncertainty, taken to be $\sigma_\beta = 1\%$~\cite{COHERENT:2024axu}.

To conclude this subsection, let us note that, in principle, a two-dimensional analysis combining both energy and timing information of the COHERENT-Ge data would be preferable. However, Ref.~\cite{COHERENT:2026yje} provides the experimental information only after projection onto either the energy or timing axis, which prevents such a two-dimensional fit. We therefore restrict our analysis to the energy spectrum and leave an update of the current analysis with inclusion of timing for future work, once the correlated dataset is made publicly available. Let us note, however, that while a timing-only analysis would help reducing the background component, it effectively reduces the information content to a single-bin-like measurement in recoil energy, thereby washing out the spectral information of the signal. This loss is particularly relevant for the determination of quantities such as the Lindhard quenching parameter $k$ and the nuclear  rms radius, whose effects are primarily induced by distortions of the recoil-energy spectrum.

\subsection{CONUS+}
\label{subsec:CONUS+}

The first run~\cite{Ackermann:2025obx} of the CONUS+ experiment took place from November 2023 to July 2024, using four high-purity Ge detectors with a total active mass of 2.83~kg, located at a baseline of $L=20.7$~m from the Leibstadt nuclear power plant in Switzerland. During the reactor-on period, a total exposure of $327~\mathrm{kg}\times\mathrm{d}$ was accumulated, distributed among three sub-detectors, while the reactor-off period corresponds to $60~\mathrm{kg}\times\mathrm{d}$. For this configuration, the reactor thermal power of 3.6~GW and the detector baseline imply an antineutrino flux of $\Phi = 1.5 \times 10^{13}~\mathrm{s^{-1}\,cm^{-2}}$. Following Ref.~\cite{Ackermann:2025obx}, we adopt $\sigma_\mathrm{noise}=48~\mathrm{eV_{ee}}$ and a Fano factor $\mathcal{F}=0.1096$.

For the analysis we follow the procedure detailed in a previous work by a subset of the present authors~\cite{DeRomeri:2025csu}. The CONUS+ results are provided in terms of excess counts, defined as the difference between the reactor-on data and the expected background model in each ionization-energy bin. This is implemented by assuming a single effective detector with threshold $160~\mathrm{eV_{ee}}$ and an effective exposure of $\mathcal{E}=119~\mathrm{kg}\times\mathrm{d}$.
Our theoretical predictions are then compared to the measured excess counts, denoted by $R_i^{\rm exp}$, through the $\chi^2$ function
\begin{equation}
\label{eq:chi2}
 \chi^2 (\vec{S}) =  \sum_i \frac{\left[R^{\rm exp}_i- (1 + \alpha)R_i^{\rm th}(\vec{S})\right]^2}{\sigma_i^2} + \left(\frac{\alpha}{\sigma_\alpha}\right)^2\, ,
\end{equation}
where $\sigma_i$ are the experimental uncertainties in each bin~\cite{Ackermann:2025obx}. The nuisance parameter $\alpha$ accounts for the overall normalization uncertainty, with $\sigma_\alpha = 16.9\%$~\cite{Ackermann:2025obx}. This uncertainty includes contributions from the reactor flux ($4.6\%$), quenching factor ($7.3\%$), energy threshold ($14.1\%$), active mass ($1.1\%$), trigger efficiency ($0.7\%$), and the weak mixing angle ($3.2\%$). The parameter vector $\vec{S}$ denotes the set of physics parameters varied in the fit, as above in the case of COHERENT.

Before closing this section, we briefly comment on other reactor-based \cevns~measurements using germanium detectors. The NCC-1701 experiment at the Dresden-II reactor reported evidence for \cevns~in 2022~\cite{Colaresi:2022obx}, though favoring a different quenching factor model, derived from iron-filtered monochromatic neutrons~\cite{Collar:2021fcl}. However, as pointed out in Ref.~\cite{Li:2025pfw}, no single choice of quenching factor can simultaneously reconcile the COHERENT-Ge, CONUS+, and Dresden-II datasets. For this reason, we decide not to include the Dresden-II result in the present analysis.
On the other hand, the TEXONO collaboration has reported limits on the \cevns~cross section using a germanium detector at the Kuo-Sheng Reactor Neutrino Laboratory~\cite{TEXONO:2024vfk}, with a total exposure of $242~\mathrm{kg}\times\mathrm{day}$. Since no \cevns~observation has been reported by TEXONO, we expect the combined sensitivity to be dominated by the higher-statistics COHERENT-Ge measurement (as confirmed by previous studies~\cite{AtzoriCorona:2025xgj,AtzoriCorona:2026wbu}), and hence we also decide not to include TEXONO in our analysis.

\section{Results}
\label{sec:results}

\begin{figure}[t!]
    \centering
    \includegraphics[width=0.48 \textwidth]{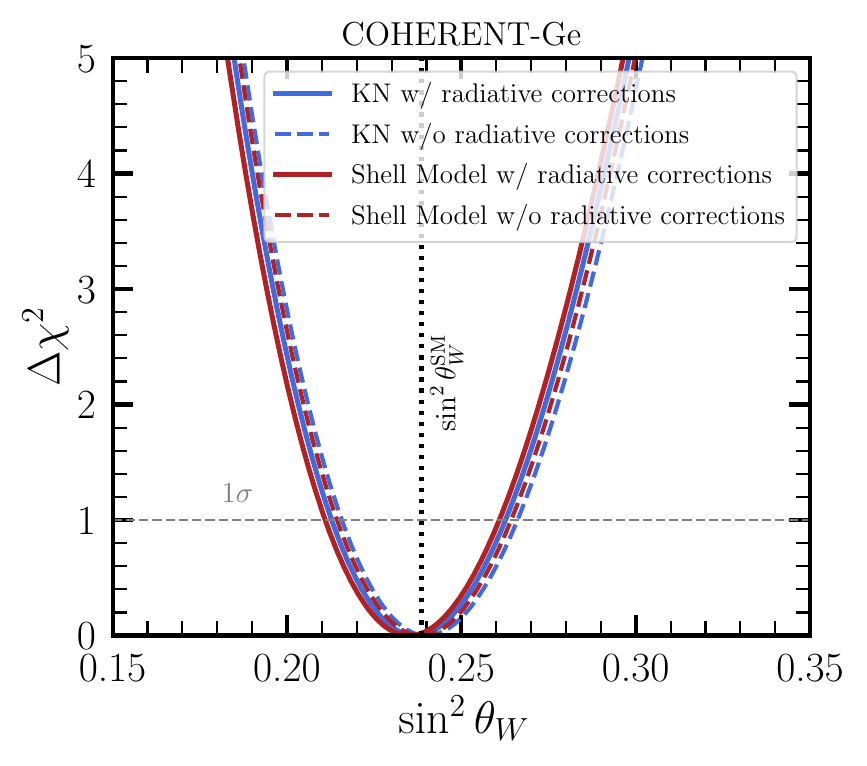}
    \includegraphics[width=0.48 \textwidth]{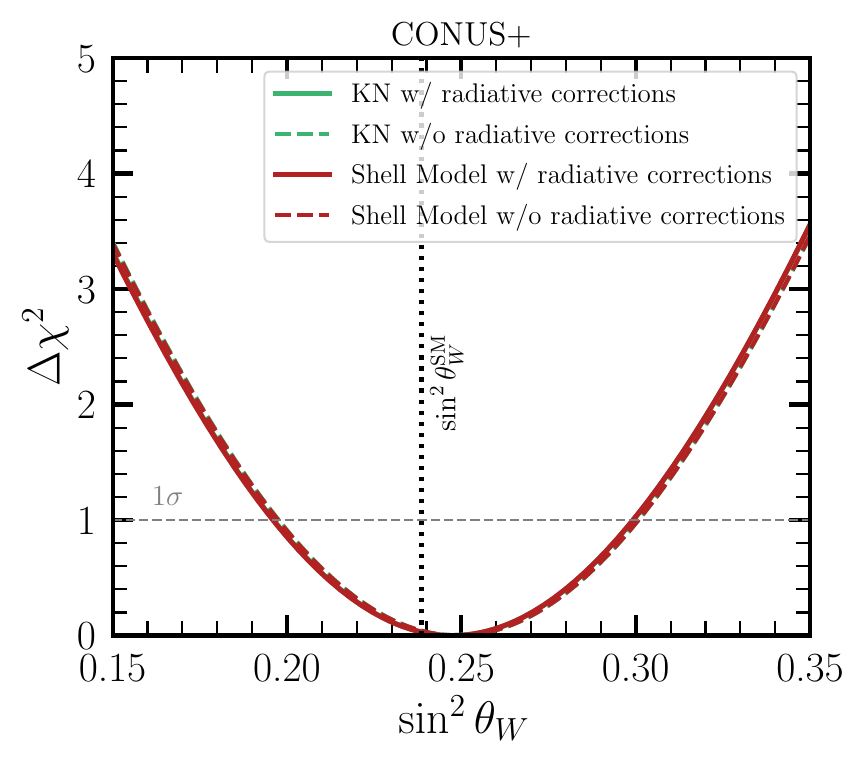}
    \caption{Reduced $\chi^2$ profiles for the weak mixing angle extracted from COHERENT Ge-Mini (\textbf{left panel}) and CONUS+ (\textbf{right panel}) data. Solid (dashed) curves correspond to calculations including (neglecting) radiative corrections to the vector couplings. In the left (right) panel, blue (green) and red (red) curves denote the KN and Shell Model nuclear form factors, respectively.   The profiles are obtained fixing the Lindhard parameter $k = 0.157$ ($k = 0.162$), for the case of COHERENT-Ge (CONUS+). }
    \label{fig:sw2}
\end{figure}

In this section, we present the results obtained for the different electroweak and nuclear physics parameters.
In an effort to obtain the most accurate theoretical predictions, we perform our calculations under four different assumptions. First, we investigate the impact of the vector couplings $g_V^p$ and $g_V^n$ on the predicted event rates by comparing tree-level and radiatively corrected cases. Second, we explore nuclear-structure effects by contrasting the phenomenological KN form factors with full nuclear Shell Model calculations. 
As anticipated, for the KN parametrization, we use the proton form factor, $F_p$, fixing the proton rms radius $\langle R_p \rangle = 4.078~\mathrm{fm}$~\cite{Wang:2024ste}, and the neutron form factor, $F_n$, with $R_A = 5.006~\mathrm{fm}$, corresponding to a neutron rms radius of $\langle R_n \rangle = 4.099~\mathrm{fm}$. In the Shell Model case, we employ the full expression given in Eq.~(58) of Ref.~\cite{Hoferichter:2020osn}, including all relevant nuclear response operators.

\paragraph{Total predicted number of events: the role of subleading effects and of nuclear models.}
Assuming tree-level vector couplings, our calculations predict 123 and 124 total \cevns~events at COHERENT-Ge for the KN and Shell Model form factors, respectively\footnote{For comparison, neglecting the $\Phi''$ contribution and momentum-dependent corrections in the $\mathcal{M}$ response yields 125 events.}. By including radiative corrections in the definitions of $g_V^p$ and $g_V^n$ these predictions increase to 126 and 127 \cevns~events.
For CONUS+ both nuclear-structure effects and radiative corrections are negligible, as expected from the much smaller momentum transfer probed in reactor-based \cevns~measurements, yielding 151 events at tree level and 152 events with radiative corrections, assuming a single detector with $160~\mathrm{eV_{ee}}$ threshold and an exposure of $\mathcal{E}=1~\mathrm{kg}\times\mathrm{d}$.

Aiming at a precise determination of these parameters, we have also verified that the axial-vector SM contribution is negligible in both experiments. Indeed, using Eq.~(62) of Ref.~\cite{Hoferichter:2020osn} and considering one-body hadronic currents, we find 0.005 total axial events for COHERENT-Ge and 0.03 for CONUS+, noting that only the $^{73}$Ge ($9/2^+$) isotope contributes {with a suppressed abundance of about $7.7 \%$ in natural Ge\footnote{For an order-of-magnitude estimate of the axial-to-vector ratio per isotope, see Eq.~(8) of Ref.~\cite{AristizabalSierra:2026rlo}.}. Before proceeding to the actual determination of the electroweak and nuclear parameters, we note that these numbers are obtained fixing the Lindhard parameter to $k=0.157$~\cite{COHERENT:2026yje} for COHERENT-Ge and $k=0.162$~\cite{Bonhomme:2022lcz} for CONUS+, as adopted by the respective collaborations. The impact of varying the quenching model is investigated in detail later in this section.

\begin{figure}[t!]
    \centering
    \includegraphics[width=0.48 \textwidth]{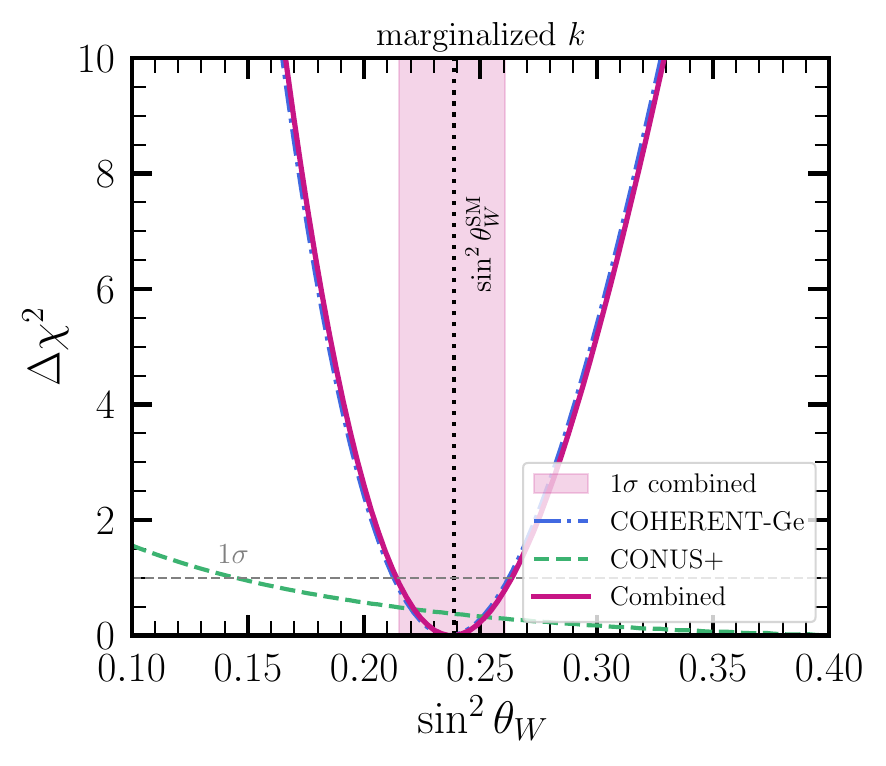}
    \includegraphics[width=0.48 \textwidth]{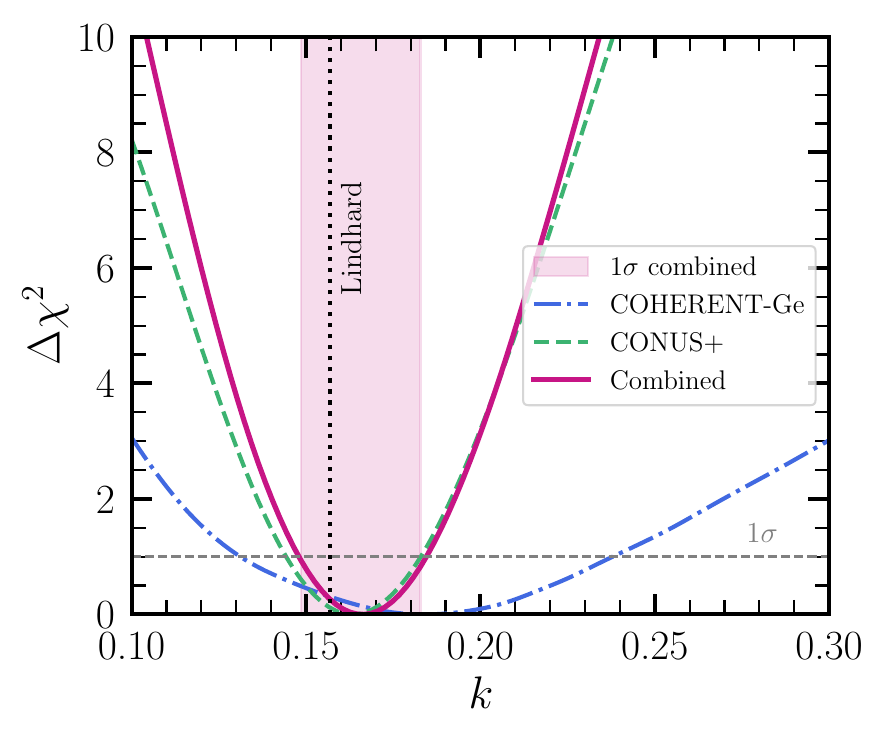}
    \caption{Reduced $\chi^2$ profiles for the weak mixing angle (\textbf{left panel}) and the Lindhard parameter (\textbf{right panel}) obtained from COHERENT Ge-Mini (blue dash-dotted), CONUS+ (green dashed), and the combined analysis of both datasets (solid magenta). The profiles for $\sin^2\theta_W$ are obtained after marginalizing over the Lindhard parameter freely, while those for $k$ are obtained fixing $\sin^2\theta_W^\mathrm{SM} = 0.23857$.}
    \label{fig:sw2b}
\end{figure}

\paragraph{Determination of the weak mixing angle.}
Figure~\ref{fig:sw2} illustrates the $\Delta\chi^2$ profiles as a function of the weak mixing angle obtained from COHERENT-Ge (left panel) and CONUS+ (right panel) analyses. In the left (right) panel, blue (green) curves correspond to the KN nuclear form factor, while red curves to nuclear Shell Model calculations. Dashed lines denote results without radiative corrections, whereas solid lines include radiative corrections in the \cevns~cross section. For reference, we also indicate with a black dotted vertical line the theoretical value of $\sin^2\theta_W^\mathrm{SM} = 0.23857$~\cite{ParticleDataGroup:2024cfk} defined in the $\overline{\text{MS}}$ renormalization scheme~\cite{Erler:2004in,Erler:2017knj}.
As can be seen from this figure, both the impact of radiative corrections and the difference between the KN and Shell Model form factor descriptions remain relatively small for the current COHERENT-Ge sensitivity. For CONUS+, these effects are even less pronounced, as expected from the lower momentum transfer involved, making the choice of nuclear model essentially irrelevant. In light of these results, the remainder of this work is based on the KN form factor together with radiatively corrected couplings.
 Before proceeding, we report the best-fit values and $1\sigma$ uncertainties for $\sin^2\theta_W$, emphasizing that, within uncertainties, the dependence on the nuclear model is negligible. For COHERENT-Ge we find 
\begin{equation}
\sin^2\theta_W =
\begin{cases}
0.240^{+0.026}_{-0.024} & \text{KN, no radiative corrections} \, ,  \\
0.236^{+0.026}_{-0.021} & \text{KN, with radiative corrections}\, , \\
0.238^{+0.026}_{-0.023} & \text{Shell Model, no radiative corrections}\, , \\
0.235^{+0.025}_{-0.024} & \text{Shell Model, with radiative corrections}\, .
\end{cases}
\end{equation}
Similarly, for CONUS+ (see also other analyses in~\cite{DeRomeri:2025csu,Alpizar-Venegas:2025wor,Chattaraj:2025fvx,AtzoriCorona:2025ygn,CONUS:2026uhz}) we find
\begin{equation}
\sin^2\theta_W =
\begin{cases}
0.249^{+0.051}_{-0.052} & \text{no radiative corrections} \, ,  \\
0.247^{+0.052}_{-0.051} & \text{with radiative corrections}\, .
\end{cases}
\end{equation}

We next assess the robustness of these results with respect to the quenching factor. The results discussed so far were obtained by fixing  $k = 0.157$ ($k = 0.162$) for the case of COHERENT-Ge (CONUS+), as adopted by the respective collaborations, which prevents a fully consistent combined fit. For the sake of comparison, the results of such an analysis are given in Appendix~\ref{sec:appendix}. We therefore repeat the analysis profiling over $k$, while removing the prior uncertainty on the quenching factor. In this case, the systematic uncertainty associated with the quenching factor is not included as an external constraint. 
The left panel of Fig.~\ref{fig:sw2b} shows the corresponding results for COHERENT-Ge (blue dash-dotted), CONUS+ (green dashed), and the combined analysis of both datasets (solid magenta). We also show as a vertical light magenta shaded band the $1\sigma$ interval obtained from the combined analysis. As clear from this plot, once $k$ is marginalized over, the sensitivity of CONUS+ to $\sin^2\theta_W$ is significantly reduced. This is expected since quenching effects strongly impact sub-keV recoil energies relevant for CONUS+, as has been  pointed out in Ref.~\cite{Li:2025pfw}. 
On the other hand, COHERENT-Ge operated with a higher effective threshold of $4~\mathrm{keV_{ee}}$, thus exhibiting a much weaker dependence on $k$. As a result, the combined constraint is driven by COHERENT-Ge and reads
\begin{equation}
\sin^2\theta_W= 0.237^{+0.027}_{-0.022},
\qquad \textrm{(KN, with radiative corrections)} \, .
\end{equation}

\begin{figure}[t!]
    \centering
    \includegraphics[width=0.5 \textwidth]{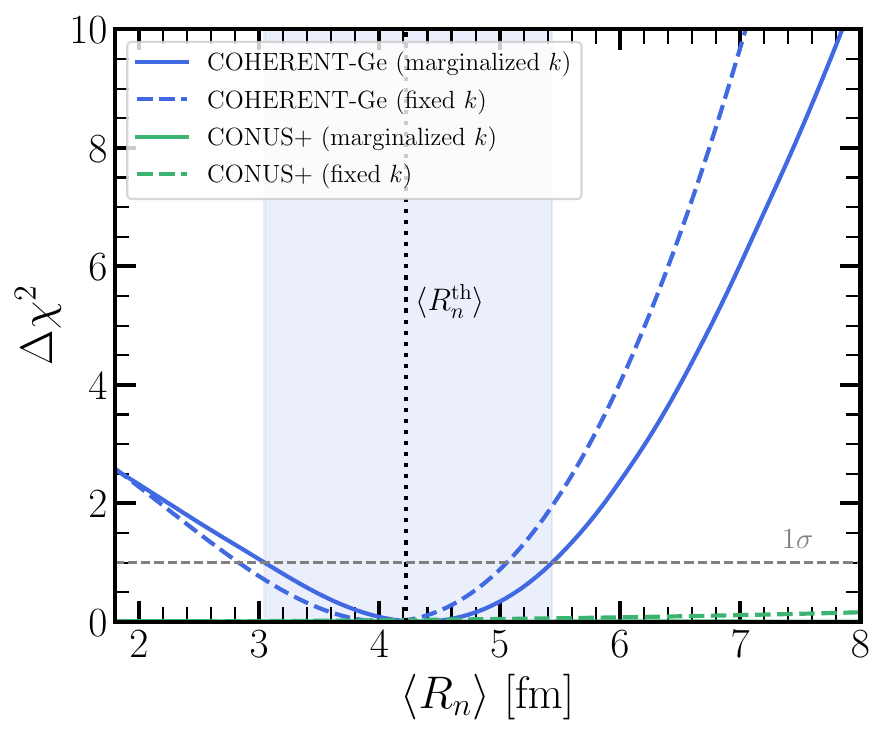}
    \caption{$\Delta\chi^2$ profiles for the neutron rms radius $\langle R_n \rangle$. Results are shown for fixed $k=0.157$ (dashed lines) and after profiling over $k$ (solid lines). COHERENT-Ge and CONUS+ results are shown in blue and green, respectively. The $1\sigma$ interval for COHERENT-Ge is also indicated.}
    \label{fig:Rn}
\end{figure}

\paragraph{Sensitivity to the Lindhard parameter $k$.}
Next, we explore the sensitivity of both datasets to the Lindhard parameter $k$. Constraining this quantity is of particular importance given the extensive use of germanium detectors in neutrino scattering experiments, dark matter direct detection, and neutrinoless double-beta decay searches, including TEXONO~\cite{TEXONO:2006xkg}, Ricochet~\cite{Ricochet:2022qim}, NUCLEUS~\cite{NUCLEUS:2019exo}, $\nu$-GeN~\cite{nGeN:2022uje}, CDMS~\cite{CDMS:2004tet}, CoGeNT~\cite{CoGeNT:2013cba}, CDEX~\cite{CDEX:2019hzn}, EDELWEISS~\cite{EDELWEISS:2022fou}, and LEGEND~\cite{LEGEND:2021bnm}, as well as proposed \cevns~programs employing germanium detectors at future facilities such as ESS and J-PARC~\cite{Baxter:2019mcx,Collar:2025sle}. 
In the right panel of Fig.~\ref{fig:sw2b} we present the constraints on the Lindhard parameter $k$ obtained from COHERENT-Ge (blue dash-dotted), CONUS+ (green dashed), and the combined analysis of both datasets (solid magenta). For the sake of comparison, we also show the empirical value coming from Lindhard theory, $k = 0.157$, as a black dotted vertical line, and the $1\sigma$ variation obtained from the combined analysis (light magenta shaded band). For this analysis, we remove the contribution associated with the quenching factor from the overall systematic uncertainty. As expected, and opposite to the $\sin^2\theta_W$ analysis, CONUS+ provides the dominant constraint, while the inclusion of COHERENT data leads to an only modest improvement in the combined fit. The resulting constraints, obtained using the KN form factor and radiatively corrected couplings, are
\begin{equation}
k_{\rm COH\text{-}Ge} = 0.183^{+0.055}_{-0.051}, \qquad
k_{\rm CONUS+} = 0.163^{+0.020}_{-0.018}, \qquad
k_{\rm Combined} = 0.167^{+0.018}_{-0.018}\, .
\end{equation}
It is interesting to note that this result is in excellent agreement with the value of $k$ preferred by the CONUS+ measurement~\cite{Bonhomme:2022lcz}, while the COHERENT-Ge determination is less precise but still fully consistent within $1\sigma$. Concerning the latter, the energy-only analysis of the COHERENT-Ge dataset performed here seems to indicate a preference for a higher value of $k$, in contrast to the value $k= 0.133 \, Z^{2/3}A^{-1/2} \simeq 0.157$ assumed by the collaboration~\cite{COHERENT:2026yje}, and corresponding to the predicted value in the Lindhard theory~\cite{osti_4536390,PhysRev.124.128}.
A closer agreement is, however, expected once a joint energy-and-timing analysis becomes feasible with future data releases. Finally, we have verified that radiative corrections have a negligible impact on the extracted values of $k$.

\paragraph{Sensitivity to the neutron rms radius.}
Next, we investigate the sensitivity of \cevns~data to nuclear-structure effects through the extraction of the neutron rms radius $\langle R_n \rangle$ of germanium. To assess the robustness of the determination, we perform two complementary analyses. First, we perform a statistical analysis fixing the Lindhard parameter to $k= 0.157$ (as predicted by the theory), and then we marginalize over $k$, as we did above in the determination of $\sin^2\theta_W$. In the first case, we remove the contribution associated with the nuclear-radius  from the systematic uncertainty, while in the second case we additionally remove the quenching factor uncertainty.
The corresponding constraints from the COHERENT-Ge dataset are shown in Fig.~\ref{fig:Rn}, with solid and dashed curves corresponding to the analyses with fixed and marginalized $k$, respectively.  As expected, CONUS+ exhibits essentially no sensitivity to $\langle R_n \rangle$ due to the very low momentum transfer involved.
The resulting best-fit values and $1\sigma$ intervals for the neutron rms radius and neutron skin are
\begin{align}
\langle R_n \rangle_{\rm fixed,~k}
&=
(4.05^{+1.01}_{-1.19})~{\rm fm}\, ,
&
\Delta_{np}^{\rm fixed,~k}
&=
(-0.02^{+1.01}_{-1.19})~{\rm fm}\, ,
\\[1ex]
\langle R_n \rangle_{\rm marg,~k}
&=
(4.37^{+1.07}_{-1.32})~{\rm fm}\, ,
&
\Delta_{np}^{\rm marg,~k}
&=
(0.29^{+1.07}_{-1.32})~{\rm fm}\, .
\end{align}
A non-negative neutron skin is generally expected on physical grounds, since in medium and heavy nuclei the neutron distribution typically extends further than the proton distribution. In this respect, the positive central value obtained after profiling over $k$ appears more physically plausible, highlighting the importance of treating quenching and nuclear-structure effects simultaneously in a careful and robust statistical analysis.

\paragraph{Two-dimensional fits.}
The correlations among the fitted parameters can be better understood from the two-dimensional contours shown in Fig.~\ref{fig:Lindhard_k_vs_Rn}. The left panel displays the allowed regions in the $\sin^2\theta_W$-$k$ plane. As anticipated from the one-dimensional profiles, the determination of $\sin^2\theta_W$ from COHERENT-Ge data (blue) is largely insensitive to the value of $k$, reflecting the relatively high recoil-energy threshold of the experiment. On the other hand, the much lower threshold achieved by CONUS+ enhances its sensitivity to low-energy spectral distortions induced by quenching effects, leading to a strong correlation between $\sin^2\theta_W$ and $k$ and preventing a simultaneous determination of both parameters.  The power of the combined fit becomes evident in the combined fit (magenta), where the allowed parameter space is significantly reduced, compared to the individual CONUS+ (green) and COHERENT-Ge (blue) regions.

\begin{figure}[t!]
    \centering
    \includegraphics[width= \textwidth]{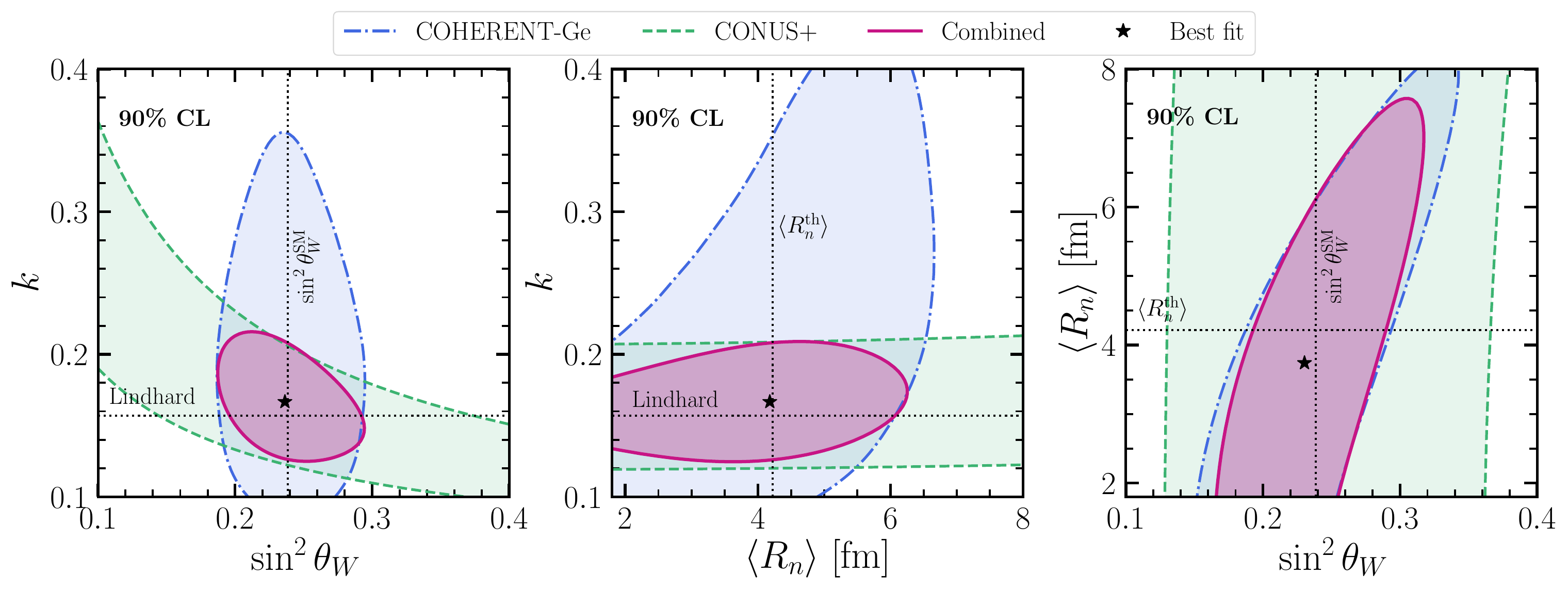}
    \caption{Two-dimensional contours showing the correlations among $\sin^2 \theta_W$, the Lindhard parameter $k$, and the nuclear rms radius $\langle R_n \rangle$ for the individual analysis of COHERENT-Ge and CONUS+ data and the combination of both.}
    \label{fig:Lindhard_k_vs_Rn}
\end{figure}

The central panel shows the correlation between the Lindhard parameter $k$ and the neutron rms radius $\langle R_n \rangle$. In this case, COHERENT-Ge exhibits a sizable correlation between the two quantities because both of them affect the shape of the recoil spectrum. The CONUS+ determination of $k$ is instead essentially insensitive to $\langle R_n \rangle$, as expected from the very low momentum transfer characteristic of reactor neutrinos. Once again, when combining the data sets the allowed region of parameter space is significantly reduced with respect to the individual analyses due to the complementarity of the experiments.

Finally, the right panel shows the contours in the $\sin^2\theta_W$-$\langle R_n \rangle$ plane, for which the sensitivity is dominated by COHERENT-Ge. The inclusion of CONUS+ only leads to a mild improvement, mainly through its ability to constrain the quenching parameter. This is expected, since CONUS+ is not sensitive to $\langle R_n\rangle$, while also not measuring $\sin^2 \theta_W$ more precisely than COHERENT.

\paragraph{Implications of different nuclear models.}
In addition to the nuclear Shell Model, alternative nuclear structure models such as those based on Skyrme-Hartree-Fock calculations in Ref.~~\cite{AbdelKhaleq:2024hir} could also be used. In that study, the nuclear form factors associated with the dominant \cevns~contribution from the Coulomb operator $\mathcal{M}$ are found to be in good agreement with the Shell Model predictions, especially in the region of interest of COHERENT-Ge. More noticeable differences appear in the subleading $\Phi''$ response, which also contributes to the vector channel; however, as discussed above, its impact on the total \cevns~rate is negligible. Similarly, for the case of the axial-vector contribution, these two nuclear models yield different predictions for the $\Sigma''$ structure functions, but this contribution is suppressed relative to the dominant vector channel and therefore does not affect our results.

\section{Conclusions}
\label{sec:concl}

The COHERENT collaboration has recently reported the most precise measurement of the \cevns~cross section to date, using their Ge-Mini germanium detector array at the Spallation Neutron Source. The measurement is in excellent agreement with the SM prediction and opens the door to a variety of phenomenological applications. In particular, its combination with the previous CONUS+ measurement of \cevns~on germanium, performed using reactor antineutrinos in a very different kinematic regime, provides a unique opportunity to exploit the complementarity between distinct neutrino sources and momentum-transfer scales.

In this work, we have presented a combined analysis of the COHERENT Ge-Mini and CONUS+ datasets, focusing on the extraction of electroweak and nuclear-structure information from germanium \cevns~data. While the low-energy reactor antineutrinos of CONUS+ probe the fully coherent regime and primarily constrain the overall normalization of the cross section, the very low threshold achieved by CONUS+ also provides enhanced sensitivity to spectral distortions at low recoil energies, particularly those induced by quenching effects. On the other hand, the higher-energy neutrinos from pion decay at rest at COHERENT-Ge probe the onset of coherence loss and are therefore directly sensitive to nuclear form factor effects. This complementarity enables a more robust disentangling of normalization uncertainties, detector-response effects, and genuine spectral distortions in the recoil-energy spectrum.

We have derived constraints on the weak mixing angle at low momentum transfer, the Lindhard quenching parameter $k$, and the neutron rms radius of germanium. In our analysis, we included radiative corrections to the vector couplings and assessed their impact on the extraction of $\sin^2\theta_W$. While these corrections are found to be somewhat more relevant for the COHERENT-Ge determination, their effect is much smaller for CONUS+, as expected from the lower momentum transfer probed by reactor antineutrinos. Furthermore, we compared predictions obtained using the phenomenological Klein-Nystrand form factor and full Shell Model calculations, finding very good agreement within the current experimental precision.

We have also explored quenching factor uncertainties. We have shown that the Lindhard parameter $k$ significantly affects the interpretation of low-energy reactor data, strongly reducing the sensitivity of CONUS+ to the weak mixing angle once marginalized over. At the same time, CONUS+ provides the dominant constraint on $k$, highlighting the advantage of combining the two experiments. We further demonstrated that both the extractions of the neutron rms radius and of  $\sin^2\theta_W$ are sensitive to the treatment of quenching effects, especially in CONUS+, emphasizing the importance of performing a consistent two-dimensional analysis of detector and nuclear-structure uncertainties.

Overall, our results show the potential of combining \cevns~measurements from different neutrino sources and detector configurations to simultaneously probe electroweak physics, detector response, and nuclear structure. Future improvements in  quenching factor measurements, increase of statistics, and multidimensional timing-and-energy analyses will further strengthen the role of \cevns~experiments as precision probes of the SM.

\vspace{1cm}

\paragraph*{Final note.} 
Upon completion of this work, two related articles appeared on  arXiv~\cite{AtzoriCorona:2026wbu,CONUS:2026uhz}. Reference~\cite{AtzoriCorona:2026wbu} performs a similar global analysis of the COHERENT-Ge, CONUS+, and TEXONO datasets, and our findings are in reasonable agreement despite differences in methodology and analysis strategy. 
In contrast, Ref.~\cite{CONUS:2026uhz} presents phenomenological implications of CONUS+ data, which are, however, not sensitive to $\langle R_n\rangle$, also covered here. In this work, we therefore complement these analyses by systematically exploring the role of quenching effects and by performing a combined extraction of nuclear-structure information where experimentally feasible.

\section*{Acknowledgments}
 V.D.R. acknowledges financial support by the grant CIDEXG/2022/20 (from Generalitat Valenciana) and by the Spanish grants CNS2023-144124 (MCIN/AEI/10.13039/501100011033 and “Next Generation EU”/PRTR), PID2023-147306NB-I00, and CEX2023-001292-S (MCIU/AEI/10.13039/501100011033). D.K.P. acknowledges funding from the European Union’s Horizon Europe research and innovation programme under the Marie Skłodowska‑Curie Actions grant agreement No.~101198541 (neutrinoSPHERE).  G.S.G. has been supported by Sistema Nacional de Investigadoras e Investigadores (SNII, México) and by SECIHTI Project No. CBF-2025-I-1589.

\appendix

\section{Results for fixed $k$}
\label{sec:appendix}

In this appendix, we present the reduced $\chi^2$ profiles for the determination of the weak mixing angle assuming different values of the Lindhard parameter $k$. For each experiment, we show in Fig.~\ref{fig:sw2_fixedk} the results obtained by fixing $k=0.157$, following the COHERENT recommendation (solid lines), and $k=0.162$, as adopted by CONUS+ (dashed lines). The corresponding combined-analysis results are also displayed in each case.
As expected, the COHERENT-Ge determination of $\sin^2\theta_W$ is essentially unaffected by the choice of $k$. In contrast, the CONUS+ analysis shows a noticeable dependence on the assumed Lindhard parameter. These results are fully consistent with the discussion of Fig.~\ref{fig:sw2b} and the left panel of Fig.~\ref{fig:Lindhard_k_vs_Rn}.

\begin{figure}[t!]
    \centering
        \includegraphics[width=0.48 \textwidth]{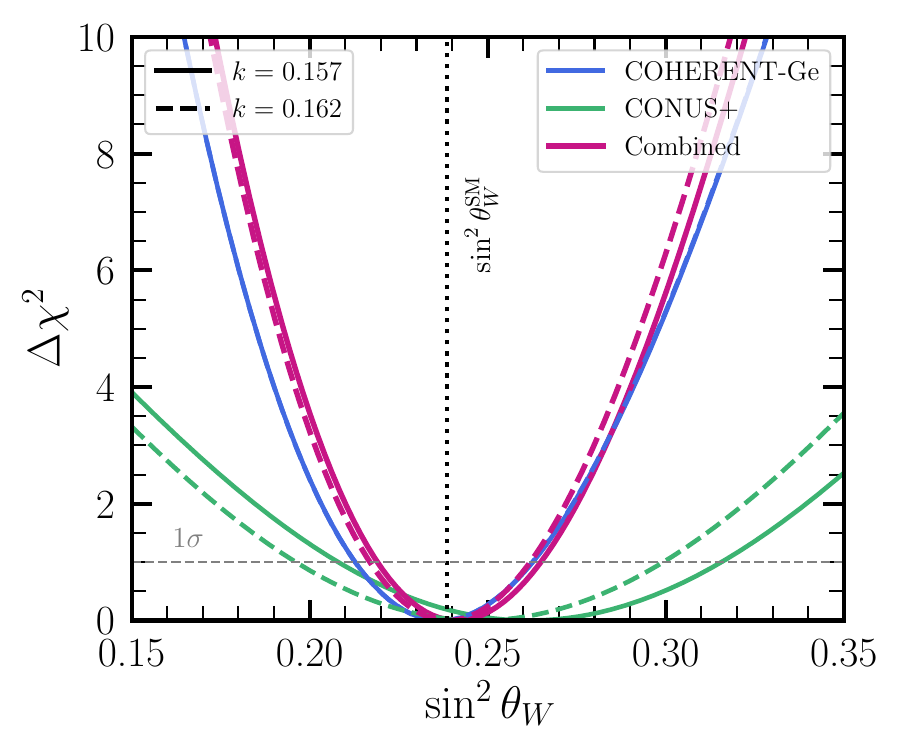}
    \caption{Reduced $\chi^2$ profiles for the weak mixing angle, fixing the Lindhard parameter $k=0.157$ (solid lines) and $k=0.162$ (dashed lines). The results are given for the individual (blue and green) and combined (magenta) analysis of COHERENT-Ge and CONUS+ datasets.}
    \label{fig:sw2_fixedk}
\end{figure}

\bibliographystyle{utphys}
\bibliography{bibliography}  

\end{document}